# A review of data-driven short-term voltage stability assessment of power systems: Concept, principle, and challenges

Jiting Cao [a], Meng Zhang [b], Yang Li [a,*]

[a] School of Electrical Engineering, Northeast Electric Power University, Jilin 132012, China

[b] National Key Laboratory of Science and Technology on Vessel Integrated Power System, Naval University of Engineering, Wuhan 430033, China

* Corresponding author. E-mail address: liyang@neepu.edu.cn (Y. Li).

**Abstract:** With the rapid growth of power market reform and power demand, the power transmission capacity of a power grid is approaching its limit, and the secure and stable operation of power systems becomes increasingly important. In particular, in modern power grids, the proportion of dynamic loads with fast recovery characteristics such as air conditioners, refrigerators, and industrial motors is increasing. As well as the increasing proportion of different forms of renewable energy in power systems. Therefore, short-term voltage stability (STVS) of power systems cannot be ignored. This article comprehensively sorts out the STVS problems of power systems from the perspective of data-driven methods and discusses existing challenges.

## 1 Introduction

Short-term voltage stability assessment (STVSA) is the linchpin for ensuring the secure and stable operation of a power system [1]. In urban load centers, the proportion of dynamic loads with fast recovery characteristics such as air conditioners, refrigerators, and industrial motors is increasing, and short-term voltage stability (STVS) problems of power systems are becoming growingly prominent. Previous works show that several serious blackouts, such as the blackouts that occurred in the northeastern United States and Canada on August 14, 2003, and in Athens, Greece on July 12, 2004, are closely related to the sharp drop of the voltage [2, 3]. In addition, due to the uncertainty of renewable energy and the increasing permeability of different forms of renewable energy, the randomness and volatility of the system operating conditions make the secure and stable operation of power systems face huge challenges. On September 28, 2016, a large number of wind turbines were disconnected from the South Australian power grid, which caused a blackout [4]. With the rapid development of power systems, the integration of large-scale renewable energy and high-proportion induction motor loads have aggravated the STVS problems.

Due to strong modeling and decision-making capabilities, machine learning techniques are playing an increasingly important role in addressing complex engineering problems [5-7]. Meanwhile, wide-area measurement protection and control (WAMPAC) has spread rapidly. By using massive phasor measurement unit (PMU) data as the information source, data-driven methods provide new opportunities for solving traditional problems in power system stability. In other words, a STVSA model can be built by leveraging measurement data and machine learning to ensure the secure and stable operation of power systems.

For the introduction and analysis of short-term voltage stability assessment, this paper surveys the relevant literature. The search for related papers is conducted using the Web of Science database. Search for "short" or "transient" in the topic. The main documents of STVSA from 2000 to 2021 are drawn, as shown in Figure 1.

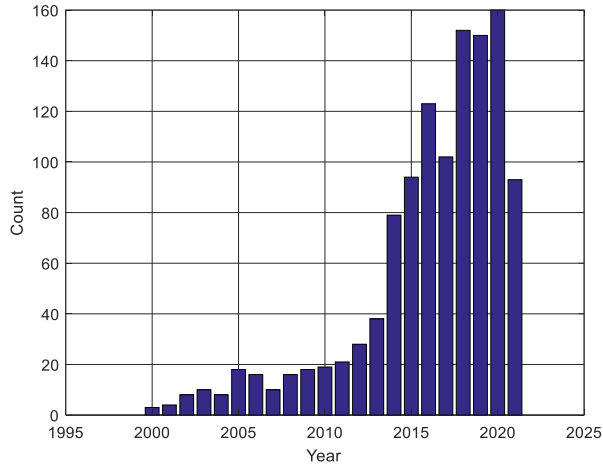

Figure 1 Number of publications per year from 2000 to 2021

It can be seen from Figure 1 that the STVS problems have attracted attention since 2000, and the number of publications per year from 2017 to 2020 is more than 100. Among them, we select the 10 journals with the most publications after 2000 and sort them by number, as shown in Figure 2.

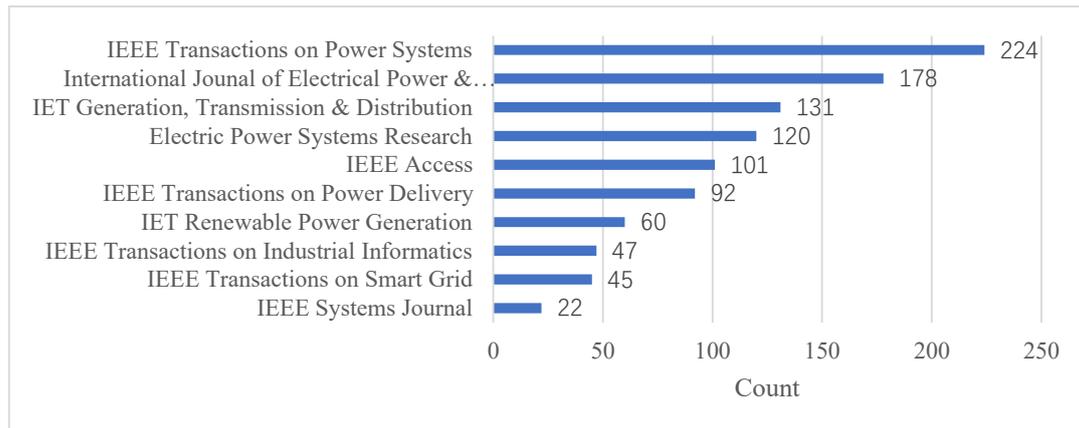

Figure 2 Summary of the number of cited journal articles

It can be seen from Figure 2 that IEEE Transactions on Power Systems and International Journal of Electrical Power & Energy Systems, respectively, are the most popular journals in the field, followed by IET Generation, Transmission & Distribution.

The purposes of this article are to organize existing data-driven STVSA methods and summarize existing challenges in this direction. The rest of this paper is organized as follows. Section 2 mainly introduces the mechanism of STVS and the classification of assessment methods. Section 3 focuses on the principle of STVSA. Section 4 lists the challenges of data-driven STVSA methods and section 5 summarizes the full text.

## 2 Concept of STVSA

### 2.1 Definition and mechanism of STVS

Power system voltage stability is mainly divided into steady-state voltage stability, STVS, and mid-term and long-term voltage stability. STVS refers to the ability of power systems to quickly restore their bus voltage to an acceptable level after a large disturbance [8]. STVS is related to the dynamic characteristics of load components with fast recovery characteristics, such as induction motors [9], power electronic loads [10], and high voltage direct current (HVDC) converters [11]. These loads tend to recover power consumption and draw excessive current after a large disturbance, which may cause short-term voltage instability (STVI) [12]. In addition, STVS problems caused by the stalling of single-phase air-conditioning compressors [13], substation failure [14], and increase of distributed energy sources [15] have also been widely concerned.

The mechanism of STVS is briefly summarized as follows. After the system is seriously disturbed, a multi-phase fault near the load center will decelerate the induction motor load. After the fault is cleared, the motor will draw a high current while trying to accelerate again. If the power system is weak, the motor may stall, which may lead to large-scale load loss and even voltage collapse [16].

### 2.2 Classification of STVSA Methods

STVSA methods can generally be divided into two types: model-based and data-driven methods.

**(1) Model-based STVSA methods**

Most previous researches on STVS focused on the dynamics of induction motor loads [17, 18]. Traditional research started from the physical mechanism, mainly including numerical integration methods based on mathematical modeling and direct methods for analyzing system energy conversion [19]. For example, reference [20] studied the interaction between induction motor load and system voltage dip. Reference [21] used the P-V curve to analyze the dynamic change process of the voltage. Reference [22] considered the dynamic characteristics of the load and judged STVS based on an energy function through a singular surface and critical fault removal time. Bifurcation theory can accurately describe the dynamic voltage change process of the system and has important practical value for the study of STVS [23]. It is also a common method to construct a complete dynamic model through the data from Supervisory Control and Data Acquisition (SCADA) and to analyze the model through time-domain simulation (TDS) [24]. In this process, the accuracy of the model is the key to solving the STVS problem [25]. Reference [26] proposed a simple method for modeling the load effect of stalled motors for STVS analysis. Reference [27] considered the electromechanical transient model of induction motors to analyze the STVS problems. Reference [28] discussed several aspects of STVI: modeling of transient overexcitation limiter for synchronous generators, possible consequences of instability, and the use of eigenvectors to identify unstable modes. Reference [29] proposed a load stability index based on load model parameters. After the load model parameters are determined, load stability can be analyzed. However, the proportion of dynamic loads and renewable energy in modern power systems is constantly increasing. The grid structure is growingly complex and the degree of non-linearity has greatly increased. There are still many difficulties in accurate physical modeling of dynamic loads [30]. This makes it difficult to perform a reliable and accurate online assessment of STVS states, which limits the effectiveness of the above model-based STVSA methods in real-time applications.

**(2) Data-driven STVSA methods**

With the rise of WAMPAC, a large amount of high-precision synchronous measurement data obtained by PMU have made data-driven STVSA methods attract widespread attention [31]. Unlike model-based STVSA methods, which rely heavily on the accuracy of physical models and model parameters, data-driven STVSA methods can get rid of this dependence and only use relevant data to assess stability. In addition, the transient time of STVS is very short, it is difficult for model-based methods to assess STVS in real-time [32], while data-driven methods can realize real-time STVSA through offline training. Reference [33] proposed a STVSA method based on an integrated graph metric set and an artificial neural network. Reference [34] developed an efficient time series (TS) data-driven scheme and a hierarchical clustering method to improve the speed of STVSA. Reference [35] proposed a pre-fault STVS prediction method based on support vector machines. Reference [36] developed an adaptive decision model based on ensemble learning to quickly predict the severity of delayed voltage recovery after system disturbances. Reference [37] proposed a hierarchical data-driven method to improve STVS by predicting load shedding. Reference [38] developed a model-free method for power system STVS monitoring, in which Lyapunov exponent (LE) was used as a proof of stability. The TS data of the PMU is used to calculate the LE to predict voltage stability in real-time. Based on reference [38], reference [39] proposed a novel framework for real-time synchrophasor data stream processing to detect system STVS by considering the actual problem of measurement noises. The framework could predict unstable conditions faster than similar existing real-time assessment methods, so it could provide more time to implement appropriate remedial measures. Reference [40] introduced a real-time STVS monitoring method driven by PMU data to avoid modeling uncertainty. At the same time, the method analyzed the impact of voltage amplitude oscillation on LE and proposed to eliminate the negative impact of oscillation the phase correction method. Reference [41] reported a model-free STVS monitor using PMU data, which used LE as a stability metric. Reference [42] proposed a new data-driven STVSA method for modern power systems and introduced a set of STVS indicators based on PMU measurement. The data-driven STVSA methods are based on real-time measurement data to predict the stability of systems, which are more suitable for online assessment of large-scale power systems.

## 3 Principle of data-driven STVSA

This section introduces the principle of data-driven STVSA in detail, which mainly includes the following: samples acquisition, feature selection, model construction, practical criteria, and indicators of STVSA.

### 3.1 Samples acquisition

Acquisition of samples is a prerequisite for assessing STVS in a data-driven method. The sources of data sets of existing researches are mainly simulation data of simulation software. The advantage of this method is that various failure scenarios and data can be customized according to research needs. Commonly used power system simulation software includes PSD-BPA, power system simulator for engineering (PSS/E), and power system analysis software package (PSASP).

PSD-BPA is introduced from the Bonneville Power Administration (BPA) by the China Electric Power Research Institute (CEPRI). Recently, it has been widely used in the field of STVSA. References [43, 44] set different fault occurrence situations in PSD-BPA. The modeling and simulations in [45, 46, 47] were carried out using PSD-BPA. By setting system load levels, dynamic load rates, and fault clearing time with different probabilities, the system operating status under various fault scenarios were simulated respectively, and then training data sets were obtained.

PSS/E is a large-scale power system simulation program developed by Protein Technologies, Inc. Corporation of the United States. It is widely used in load modeling in this field. Reference [48] set dynamic load parameters in PSS/E and generated data samples through optimal power flow calculation and TDS. Reference [49] used PSS/E for load modeling while using transient stability assessment tools for simulation. Reference [50] used PSS/E to model various loads of power systems. Reference [51] adopted PSS/E to realize the modeling of composite load and the simulation of the system.

PSASP is a comprehensive power system simulation program independently developed by CEPRI. In this field, some studies apply it to simulations of power systems to obtain data samples. By setting different parameters in PSASP, such as different load levels, motor load rates, and fault removal time, the actual operating state of the power system is simulated [52, 53]. Reference [54] used PSASP to perform TDS to obtain a transient curve of the power system and identify generator set at a specific fault.

### 3.2 Feature extraction and feature selection of STVSA

In recent decades, the rapid development of WAMPAC has greatly enhanced the situational awareness of modern power systems. In this case, geographically dispersed PMUs can simultaneously obtain a large amount of high-precision data, which brings huge benefits to data-driven STVSA. However, how to extract spatio-temporal features from real-time data that can characterize the dynamic trajectory of the system is still a key issue.

Considering limited feature learning capabilities and data fitting capabilities of shallow machine learning, it is necessary to perform further feature extraction and feature selection on original data input to the assessment model to achieve better learning results. References [52, 53] used TSSC to mine TS data features. TSSC aims to capture the key evolutionary trends and characteristics from the fluctuation transients of TS, and the learning results show the advantages of TSSC in solving STVS problems. However, due to the cumbersomeness of TS data, there is a heavy computational burden in the process of extracting shapelets. Considering that the amount of PMU data in a large-scale power system might be very large, reference [55] extracted data features based on a symbolic aggregate approximation method and developed an efficient and lightweight machine learning for STVSA.

However, although the above-mentioned methods can excavate features well, they cannot fully capture the potential time dependence contained therein. To solve this problem, reference [43] proposed BiGRU with attention mechanism, which can bi-directionally learn TS features, and automatically assign attention weight. Reference [56] built a STVSA model based on LSTM to fully capture the potential time correlations from the dynamic trajectory of systems after a disturbance. In data-driven STVSA researches, the spatial correlations of system stability states should also be considered. When subjected to a large disturbance, the weak link of the system buses may first experience STVI, and then the instability spreads to a larger area over time. Therefore, the combination of spatial correlations and temporal correlations can fully explore the key characteristics of STVS, thereby obtaining more reliable results. Reference [45] regarded LSTM as the main learning algorithm to extract time dependence. At the same time, by formulating spatial correlations as spatial attention factors to correct the sequential inputs of LSTM, the spatial correlations were further incorporated into LSTM. The correction mechanism could work together compatible with LSTM to extract and learn spatio-temporal features. Reference [50] proposed a machine learning framework that combined GCN and LSTM. Among them, GCN and LSTM were respectively used to capture spatial and temporal characteristics. This method could capture the evolution of multiple temporal and spatial trends of STVS and predict STVS results. Reference [46] proposed a spatio-temporal GCN to extract the spatio-temporal features dynamically presented after faults. Firstly, GCN was used to integrate network topology information into the learning model to utilize spatial information; then, a one-dimensional convolutional neural network (1D-CNN) was used to mine temporal correlations. In this way, the temporal and spatial characteristics of STVS were extracted through the complete convolution structure. Reference [47] proposed a sequential feature learning approach: first, based on visualized voltage contours, a comprehensive spatio-temporal sequence model was cleverly constructed to dynamically characterize the evolution trend of multiple spatio-temporal STVS. Then, the TSSC method was used to ingeniously extract the key continuous STVS features in sequence. Reference [57] used joint mutual information maximization to feature extraction and extreme gradient boosting to assess STVS of power systems with HVDC integration.

### 3.3 Model Construction of data-driven STVSA

At this stage, STVSA is facing problems such as the continuous expansion of power systems and the increase of uncertain factors. Data-driven methods can realize STVSA by fitting complex data relationships. In other words, the relationship between the STVS states and the operating conditions of power systems can be constructed using appropriate machine learning algorithms. At present, many researchers are applying advanced data-driven methods to STVSA.

#### 3.3.1 Machine learning

Before introducing data-driven STVSA methods, this article briefly introduces machine learning first. Machine learning is always regarded as an important tool for data analysis and mining in the era of big data. From the perspective of pattern recognition, STVSA can be seen as a pattern classification task. The so-called pattern recognition is to use computational methods to divide samples into certain categories according to their characteristics. The methods to solve the pattern recognition problem can be summarized into two categories: knowledge-based methods and data-based methods. The latter can be seen as a special case of data-based machine learning, where the learning goal is discrete classification. The basic principle of such a machine learning system is shown in Figure 3.

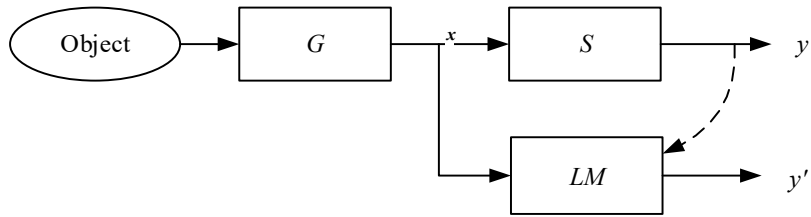

Figure 3 Data-based machine learning

In Figure 3, $G$ represents the process of observing features from an object, and feature vectors $x$ and y represent the nature of the object we care about, i.e. the classification in pattern recognition. $S$ represents the system that determines the relationship between $x$ and $y$. Data-based pattern recognition is to train the learning machine $LM$ by using such training samples and establish a mathematical model that realizes the judgment of the category y from the feature vector $x$ to calculate (predict) the category of the unknown sample.

In general, traditional shallow machine learning structures include a single hidden layer or even no hidden layer, which are effective in solving simple or well-constrained problems, but their ability to fit complex functions is limited. When the feature dimension is too large, shallow machine learning may suffer from the curse of dimensionality, while its accuracy will be greatly reduced. When dealing with complex practical application problems, shallow machine learning has great limitations and its generalization ability is restricted [58].

In this context, deep learning solves the aforementioned problems in the following manners:

(1) Deep learning networks contain multiple hidden layers, which use multi-stage transformations to describe data features and represent low-level features as abstract high-level features [59]. Therefore, they have more excellent feature expression ability, powerful function fitting ability, and generalization ability [60].

(2) The initial parameters of deep learning models are obtained through a large number of sample data training, and feature extraction of the models is through independent learning, no longer relying on manual experience.

(3) Deep learning guarantees the integrity of information to the greatest extent and provides an effective means of expressing features for data with complex features.

### 3.3.2 Data-driven STVSA based on shallow machine learning

The applications of data-driven STVSA field were initially mainly reflected in the widespread use of shallow machine learning. Shallow machine learning, which mines characteristic information of real-time measurement data, establishes mappings from the dynamic information of input systems to stable/instability states of output and finally assesses STVS. Reference [61] constructed a training model based on random forest (RF) for real-time STVSA. Reference [48] proposed a hierarchical intelligent system assessment method based on neural networks with random weights (NNRW). Reference [52] combined with a decision tree (DT) and proposed an online STVSA method based on time series shapelet classification (TSSC). Based on DT and fuzzy rules, reference [55] used a light TS learning machine to assess STVS. Reference [53] used DT to implement the STVSA process for imbalanced learning problems. Based on an extreme learning machine (ELM), reference [49] constructed an adaptive STVSA system. Reference [62] reported a hybrid random ensemble assessment model based on ELM and random vector functional link networks (RVFL), which could complete the STVSA process in a short time. Reference [63] introduced a new TS prediction method by introducing the least square support vector machine (LS-SVM) into online TS learning to predict the slip trajectory of induction motors. The above STVSA methods based on different shallow machine learning algorithms are compared, and the results are shown in Table 1.

Table 1 Comparison of STVSA methods based on shallow machine learning

| Reference | Shallow machine learning methods | Input features | Time adaptive framework |
|---|---|---|---|
| [61] | RF | $U$ | × |
| [48] | NNRW | $U/\theta/P/Q$ | × |
| [52] | DT | $U/I/P/Q$ | × |
| [55] | DT | $U/P/Q$ | × |
| [53] | DT | $U/I/P/Q$ | × |
| [49] | ELM | $U$ | √ |
| [62] | ELM and RVFL | $U$ | √ |
| [63] | LS-SVM | $U$ | × |

From Table 1, it can be seen that the input features mainly include voltage ($U$), phase angle ($\theta$), current ($I$), active power ($P$), and reactive power ($Q$); and that relatively few studies adopt time adaptive frameworks in the STVSA methods based on shallow machine learning.

### 3.3.3 Data-driven STVSA based on deep learning

STVSA models based on shallow machine learning have low fitting ability, which are unable to face the increasingly complex STVSA problems. Deep learning benefits from its powerful ability to extract the best features, and has a good application prospect in STVSA. Reference [43] proposed data augmentation based on least squares generative adversarial networks to solve the problem of insufficient samples, and assessed STVS through a bi-directional gated recurrent unit (BiGRU) with attention mechanism. To fully learn potential time correlations from the dynamic trajectory of power systems after a disturbance, reference [56] applied a long short-term memory (LSTM) to STVSA. Aiming at spatial and temporal correlations inherent in the complex transient process of smart grids, reference [45] proposed an intelligent machine learning method, which took the corrected system dynamic trajectory as input and used a LSTM-based algorithm to learn STVS features of sequence data, and a highly reliable and robust classification model could be obtained for online STVSA. Reference [64] proposed a two-stage TS deep learning framework by using a 1D-CNN for real-time STVSA. The first stage quickly detected voltage collapse, and the second stage quantified the severity of the STVS events. Reference [50] proposed a STVS online prediction method that combined graph convolutional networks (GCN) and LSTM, which could capture multiple spatio-temporal STVS evolution trends and predict STVS states. Reference [46] proposed a model framework containing spatio-temporal information. Based on GCN and 1D-CNN, it successfully connected temporal data and topological information to construct a higher-performance classification model. The specific comparison of the above STVSA methods based on deep learning is shown in Table 2.

Table 2 Comparison of STVSA algorithms based on deep learning

| Reference | Deep learning algorithm | Electrical feature quantity | Time adaptive framework | Temporal feature | Spatial feature |
|---|---|---|---|---|---|
| [43] | BiGRU with attention mechanism | $U/P/Q$ | × | √ | × |
| [56] | LSTM | $U/P/Q$ | × | √ | × |
| [45] | LSTM | $U/P/Q$ | √ | √ | √ |
| [64] | 1D-CNN | $U/I/P/Q/f$ | √ | √ | × |
| [50] | GCN | $U/P/Q$ | √ | √ | √ |
| [46] | GCN and 1D-CNN | $U/P/Q$ | √ | √ | √ |

From the comparison results in Table 2, it can be seen that the selection of electrical characteristic quantities is mainly $U/P/Q$, except for [64], which selects $I$ and frequency ($f$). Compared with the STVSA methods based on shallow machine learning in Table 1, the STVSA methods based on deep learning in Table 2 fully take into account the temporal and spatial characteristics and make full use of PMU measurement information.

### 3.4 Practical criteria for determining STVS

At present, there is no universally recognized reliable quantitative criterion for determining STVS, which brings a lot of difficulties to STVSA. In actual systems, some commonly used criteria for determining STVS are as follows:

(1) Western Electricity Coordinating Council

a) In the case of *N*-1 element failure, the time for the transient voltage to be lower than 80% of the rated voltage should not exceed 400ms; for *N*-2 element failure, the time for the transient voltage to be lower than 80% of the rated voltage should not exceed 800ms [65].

b) In *N*-1 contingencies, the voltage drop of the load bus should not exceed 25%, the non-load bus should not exceed 30%, and the voltage drop of the load bus should not exceed 20 cycles [66].

c) The voltage drop on any bus bar shall not exceed 30%. In the N-2 emergency, the voltage drop on the load bus bar shall not exceed 20% of 40 cycles [67].

(2) Tennessee Valley Authority

a) In *N*-1 contingencies, the transmission system voltage should be restored to 90% of the nominal system voltage within 0.5s after the fault is cleared [68].

b) The transmission system voltage should be restored to 90% of the nominal system voltage within 0.5s after the fault is cleared or the unit trips [69].

(3) State Grid Corporation of China

In the transient process after the power system is disturbed, the load bus voltage can be restored to above 0.80 pu within 10s [70].

(4) China Southern Power Grid

After the fault is removed, if the bus voltage is continuously lower than 0.75pu for more than 1s, the transient voltage will be unstable [71].

The engineering criterion based on a fixed voltage threshold that is often used in actual operation is simple and easy to implement. However, the threshold setting lacks sufficient theoretical support, and its reliability and selectivity are difficult to guarantee. At the same time, it cannot take into account the safety and economy of the system [72]. Considering this problem, reference [52] combined the existing knowledge of power system voltage stability and used constraint-partitioning K-means to judge STVS. The research results showed that this method was suitable for reliable qualitative assessment and judgment of STVS in a data-driven manner in the absence of reliable quantitative criteria.

**3.5 Indicators of STVSA**

In data-driven STVSA methods, PMU measurement data is fully utilized. These PMU data are used to calculate STVS indicators to monitor the degree of system voltage instability. For example, in references [38-40], based on PMU measurement data, LE is used as a STVS indicator to monitor STVS status. However, although the above research can determine whether the systems can maintain STVS, it cannot distinguish the STVS/STVI degree of power systems under different operating conditions.

Considering this issue, some indicators have been used to quantify STVS. In reference [73], a voltage stability index (VSRI) was used to assess STVS. However, STVS monitoring using the VSRI indicator requires a time interval of more than 10 seconds, so it is not suitable for online applications. Reference [51] proposed a transient voltage severity index (TVSI) to quantify the transient voltage performance of buses after interference is removed. Reference [48] developed a hierarchical intelligent system to predict unacceptable dynamic voltage deviation based on TVSI after determining whether the system has a voltage collapse. Reference [49] used a layered adaptive STVSA method and proposed a new STVS indicator, called root-mean-squared voltage-dip severity index, to quantitatively assess the severity of FIDVR. Reference [74] proposed a voltage sag severity index, which not only reflected the post-fault STVS of the system but also quantified the severity of FIDVR. Considering that different systems had different requirements for voltage sag management, reference [75] proposed a flexible method to establish a fuzzy voltage sag index (FVSI) that represented the relative interference level of a voltage sag event. To evaluate the influence of HVDC on STVS of a receiving end system, reference [76] proposed a quantitative evaluation index that considered fast-acting HVDC control, HVDC reactive power transients, induction motors, and transmission grid topology.

The above assessment indicators reflect and analyze STVS only from a single dimension. Reference [54] proposed a two-dimensional index to assess the STVS margin, where margin-to-stalling corresponded to the stall of the motor, and margin to critical rotor angle variation corresponded to the change of the rotor angle. Reference [77] proposed a practical continuous SVSI based on the voltage curve. The SVSI was composed of three parts, which reflect the degree of transient voltage recovery, the degree of transient voltage oscillation, and the ability to reach a steady state. This indicator gave a continuous assessment result of the STVS, which could be used to optimize off-line preventive control to improve the STVS.

## 4 Existing challenges in data-driven STVSA

### 4.1 Interpretability

At present, there are two main ways of thinking about the interpretability of machine learning models: establishing an interpretable machine learning model and interpreting a learned model. For data-driven STVSA, the built assessment model directly seeks mapping relationships between the system state parameters and stable indicators from data samples by exploiting the strong fitting ability and fast calculation speed of machine learning techniques.

Reference [78] proposed a local proxy model based on weighted linear regression and regularization to explain the original model. Based on first-order control variable sensitivity, reference [79] built a post hoc interpretability total transfer capability evaluation network. However, poor interpretability and the inability to balance generalization and interpretability are still challenges faced by data-driven STVSA methods.

### 4.2 Missing data

To obtain an assessment model with high performance, sufficient training data is critical. Since failures or interferences may occur in all links including data collection, measurement, transmission, and conversion, missing data will affect the accuracy of assessment models in the decision-making analysis process [80].

At present, some scholars have carried out preliminary research on the problem of missing data in online dynamic security assessment. To solve this problem, reference [81] proposed a complete data-driven method for pre-failure dynamic security assessment based on incomplete data measurement of PMU. In this method, a generative adversarial network is used to solve the problem of missing data. Reference [82] based on the integrated network of RVFL, proposed a data loss tolerance method for STVSA after failures.

Based on the above analysis, it can be found that the existing missing data solution approach mainly focuses on ensemble learning. However, these ensemble learning methods rely on the observability of the PMU data. Deep generative learning is a promising solution approach for addressing this issue.

### 4.3 Class imbalanced learning

Affected by factors such as network topology, component status, and fault type, STVS process scenarios show diversity. However, there are mostly stable data samples, and less unstable samples in large disturbance scenarios, which can easily lead to the problem of class imbalance. If handled improperly, samples of small categories would be ignored and mistaken as wrong, which will greatly reduce the classification performance of machine learning models. In addition, misdetection of power system instability may lead to irreversible voltage collapse or catastrophic power outages, and erroneous marking of stable conditions as unstable can usually be remedied at a much lower cost. This means that imbalanced data sets are also cost-sensitive, that is, there are very few unstable situations, but the cost is quite expensive.

Reference [53] proposed a cost-sensitive learning-based decision tree for STVSA by using the forecasting-based nonlinear synthetic minority oversampling technique for oversampling, which generated reasonable samples by predicting the upcoming operating state of the system and set different costs for stable and unstable samples to avoid misclassification of unstable samples to the greatest extent. However, load forecasting technology will inevitably bring errors to the data, and the most appropriate misclassification cost in cost-sensitive learning is difficult to determine.

### 4.4 Adaptability to topology changes

As the increasing power demand and electricity market reforms, the operating point of a power system is getting closer to the stability limit. This will cause a series of chain reactions such as overloading of adjacent components, overloading of branches, chain overloading, machine cutting, and load shedding. At this time, the topology of the system network will face tremendous changes. The problem of power grid topology has become increasingly prominent.

At present, some studies have considered the topological association between regions. However, these existing studies mainly focus on the coupling relationship within power systems and the extraction of spatial features. There is no in-depth analysis on the adaptability of data-driven STVSA methods when system topology changes significantly. Therefore, in the following research, the adaptability to topology changes can be taken into consideration to improve the applicability and practicability of data-driven methods to modern power grids.

### 4.5 Impact of renewable integration

As we all know, vigorously developing and utilizing renewable energy resources is the key to solving the energy crisis and environmental pollution problems [83, 84]. As an effective way to integrate renewable energy, microgrids and integrated energy systems (IES) have received more and more attention all over the world. Reference [85] put forward a chance-constrained programming-based scheduling approach for microgrids with renewables. References [86, 87] investigated the coordination scheduling of electric vehicle (EV) battery swapping stations and EV charging stations in renewable microgrids, respectively. Reference [88] adopted automated reinforcement learning to perform multi-period renewable forecasting, which is beneficial to improve the operational economy of microgrids. References [89, 90] studied the scheduling of IES with renewables. Reference [91] proposed a bi-level scheduling model for isolated microgrids that considered multi-stakeholders to coordinate demand response and uncertainty of renewable energies. Reference [92] improved the operating economics of IES by coordinating integrated demand response and the uncertainty of renewable energies. Reference [93] used a stochastic optimization scheduling method to solve the energy management and pricing problems of a multi-energy interactive multi-community IES, and used a generative adversarial network to generate scenarios to solve multiple uncertainties.

However, the inherent strong uncertainty and volatility of renewables together with the high impedance, low inertia characteristics of power electronic grid-connected ports have brought new challenges to the STVS of power systems [94, 95]. Reference [96] investigated the voltage supportability of PV systems for enhancing the STVS of power systems. The control of PV systems was studied for improving the STVS in reference [97]. Reference [98] studied how to improve the STVS in residential girds with rooftop PV units. Reference [99] used PV units to improve the STVS of power systems under asymmetrical faults. Reference [100] studied the STVS of transformerless low-voltage distribution networks with small-scale PV and provided countermeasures to alleviate STVI. Reference [101] proposed an approach for post-fault STVS analysis by establishing a quantitative relationship between the system with doubly-fed induction generator-based wind farms states and external disturbances. These studies demonstrate that with the increasing penetration of renewable energies, how to maintain the STVS of power systems is becoming a complicated and challenging task.

## 5 Conclusion

Due to the access of induction motors and power electronic loads with fast recovery characteristics, the STVSA of power systems becomes increasingly significant. At the same time, as the scale of power systems continues to expand, data-driven methods have attracted more and more attention. This article aims to summarize the latest researches and ideas on the data-driven STVSA method. By collecting and summarizing the references in scientific publications, the latest developments on the data-driven STVSA methods are provided. The research carried out strongly shows that data-driven methods have great potential, especially when the operating conditions of power systems become increasingly complex.

In the existing data-driven methods, shallow learning models and deep learning models account for a large proportion. Most shallow learning methods use bus voltage as the characteristic quantity, and part of the work will consider active power, reactive power, current, and voltage phase angles. Because of the limited learning ability of shallow learning machines, most jobs will first perform feature selection to obtain a learning model with good performances [102]. Deep learning methods have better feature expression ability, which can learn more feature quantities. The feature quantities in the works of literature are mostly bus voltage, active power, and reactive power, and some works of literature also consider current, frequency, and voltage phase angles. In shallow learning models and deep learning models, some references consider the construction of time-adaptive assessment models. Based on assessing whether the system is stable, different quantitative indicators are proposed to obtain the stability margin.

In future research, the above-mentioned challenges in data-drive STVSA methods can be mitigated or even solved by using novel techniques. The interpretability of data-driven methods can be improved by constructing interpretable assessment models or interpreting the assessment results of existing models [103]. The impact of data missing in actual situations can be reduced by using new data imputation techniques like super-resolution perception [32] or deep generative learning [104]. The problem of class imbalanced learning can be solved by methods such as resampling techniques and cost-sensitive learning. It is also interesting to optimize a learning model using automated machine learning [88] or heuristic optimization techniques [105, 106]. Another interesting topic is to consider the adaptability of assessment models to topology changes by learning the network topology of a power system. The authors hope that these challenges can be solved in future research so that STVSA can be carried out more quickly and effectively in real-world applications.

# Acknowledgements

This work is supported by the Natural Science Foundation of Jilin Province, China under Grant No. YDZJ202101ZYTS149.

# Conflict of interest statement

On behalf of all authors, the corresponding author states that there is no conflict of interest.

# Data Availability Statement

No data were used to support this study.